\documentstyle[12pt]{article}

\begin{document}

\begin{flushright}

IMSc/2019/03/03  

\end{flushright} 

\vspace{2mm}

\vspace{2ex}

\begin{center}

{\large \bf Non singular M theory Universe in} \\

\vspace{4ex}

{\large \bf Loop Quantum Cosmology -- inspired Models} \\



\vspace{8ex}

{\large  S. Kalyana Rama}

\vspace{3ex}

Institute of Mathematical Sciences, HBNI, C. I. T. Campus, 

\vspace{1ex}

Tharamani, CHENNAI 600 113, India. 

\vspace{2ex}

email: krama@imsc.res.in \\ 

\end{center}

\vspace{6ex}

\centerline{ABSTRACT}

\begin{quote} 

We study an M theory universe in the Loop Quantum Cosmology --
inspired models which involve a function, the choice of which
leads to a variety of evolutions. The M theory universe is
dominated by four stacks of intersecting brane--antibranes and,
in general relativity, it becomes effectively four dimensional
in future while its seven dimensional internal space reaches a
constant size. We analyse the conditions required for non
singular evolutions and obtain explicit solutions in the
simplified case of a bi--anisotropic universe and a piece--wise
linear function for which the evolutions are non singular. One
may now ask whether the physics in the Planckian regime can
enhance the internal volume to phenomenologically interesting
values. In the simplified case considered here, there is no non
trivial enhancement. We make some comments on it.

\end{quote}

\vspace{2ex}


\vspace{2ex}









\newpage

\vspace{4ex}

\begin{center}

{\bf 1. Introduction} 

\end{center}

\vspace{2ex}

The $(9 + 1)$ dimensional superstring theory, equivalently the
$(10 + 1)$ dimensional M theory, is considered to be a quantum
theory of gravity. Any candidate for a quantum theory of gravity
may be expected to provide, among other things, a detailed
description of black hole physics and also of the beginning of
the universe. For example, such a theory should explain black
hole entropy and Hawking radiation, and should resolve the black
hole and the big bang singularities which occur in general
relativity descriptions.

String/M theory has provided detailed descriptions of black hole
entropies and Hawking radiations for certain classes of extremal
and near extremal black holes. The black holes are described by
appropriate stacks of intersecting brane--antibranes, their
entropies arise from the degrees of freedom living on these
branes, and the Hawking radiation arise from interactions
between these degrees of freedom. See, for example, \cite{sv} --
\cite{b3ps}. As for the black hole or the big bang
singularities, there are no similarly detailed string/M
theoretic descriptions although there have been a variety of
ideas. See \cite{bowick} -- \cite{k10} for a sample of them.

The $(3 + 1)$ dimensional Loop quantum gravity (LQG) based on
Ashtekar variables is considered to be another candidate for a
quantum theory of gravity \cite{ashtekar} -- \cite{b4ooks}. The
areas and volumes are quantised in LQG and the black hole
entropies are described in terms of the quanta of area \cite{rs}
-- \cite{abck}. Quantising the homogeneous sector of LQG leads
to Loop quantum cosmology (LQC) and it provides a resolution of
big bang singularity : instead of ending in a big bang
singularity, the universe undergoes a bounce when its density is
Planckian. As one goes back in time, a large universe contracts
as in general relativity, then reaches a minimum size where its
density is Planckian, bounces back from this minimum, and starts
expanding again as in general relativity as one goes further
into its past \cite{b} -- \cite{status}.

The quantum evolution of a $(3 + 1)$ dimensional universe in LQC
can be described well by effective equations which reduce to
general relativity equations in the classical limit
\cite{status}. Recently, we have constructed LQC -- inspired
models by empirically generalising these effective equations to
$(d + 1)$ dimensions and studied several aspects of these models
\cite{k16, k17, k18}. These models are characterised by two
functions but we will fix one of them by working in what is
referred to as $\bar{\mu}-$scheme. The remaining function can be
chosen so as to lead to general relativity, or to LQC, or to a
variety of evolutions, singular as well as non singular. For
example, one can model a bouncing universe or an universe which
enters and stays in the `Hagedorn phase' where the density and
temperatures are constant \cite{k17}.

In string or M theory universes, the spacetime is ten or eleven
dimensional. They all have big bang singularities when evolved
using general relativity equations. These singularities may now
be resolved in the LQC -- inspired models. The string/M theory
universes may then have a bounce instead of a big bang
singularity, or a variety of more general non singular
evolutions.

In this paper, we study the evolution of a $(10 + 1)$
dimensional M theory universe in the LQC -- inspired models. We
consider the universe studied in \cite{m06} -- \cite{k10} which,
for entropic reasons, is dominated by four stacks of
intersecting brane--antibranes. In general relativity, due to
the U--duality relations among the densities and the pressures,
this universe becomes effectively $(3 + 1)$ dimensional in
future while the seven dimensional internal space reaches a
constant size \cite{k207, k08, k10}.

In the present study, we first analyse qualitatively the
conditions required for non singular evolution in the LQC --
inspired models. Then we simplify our set up in order to obtain
explicit solutions : Instead of considering a general
anisotropic universe, we consider a bi--anisotropic universe
where the space is $d = (\tilde{n} + n)$ dimensional and where
the quantities corresponding to the $\tilde{n}$ and the $n$
dimensional spaces are seperately isotropic; and, consider a
simplified, piece--wise linear function for which the evolutions
are non singular.

We obtain explicit solutions for a bi--anisotropic universe and
then consider the M theory universe for which the evolution is
non singular, and which becomes effectively $(3 + 1)$
dimensional in future while its internal space reaches a
constant size. One may now ask whether the physics in the non
singular Planckian regime can enhance the future constant size
of the internal space. Such a large internal space, obtained
with no fine tuning, may be useful in phenomenological model
building, see for example \cite{add, cq}. Answering this
question using the explicit solutions obtained in this paper, we
find no non trivial enhancement of the internal size.

Although this answer may be disappointing and is perhaps not
unexpected, we like to emphasise that it is now possible to ask
such a question and to seek its answer for an M theory universe
in the LQC -- inspired models. This is because the question
itself is meaningful, and its answers may then be sought, only
if a higher dimensional universe evolves non singularly in the
Planckian regime, and only if it dimensionally compactifies in
future. However, more analysis is needed to determine whether or
not a large volume compactification is possible in the LQC --
inspired models but this is beyond the scope of the present
paper.

The LQC -- inspired models are constructed empirically and,
hence, are limited in scope. Nevertheless, they have several
uses as toy models. In this paper, we also provide a critical
discussion of the limitations and the possible uses of these
models.

This paper is organised as follows. In section {\bf 2}, we
present the equations of motion in general relativity and in the
LQC -- inspired models. In section {\bf 3}, we present the
density and the pressures for the most entropic constituents of
an M theory universe, incorporating U--duality relations. In
section {\bf 4}, we analyse qualitatively the general evolution
and make some simplifying assumptions. In section {\bf 5}, we
obtain explicit solutions.  In section {\bf 6}, using these
solutions, we analyse the size of the internal space. In section
{\bf 7}, we discuss critically the limitations and the uses of
LQC -- inspired models. In section {\bf 8}, we summarise the
paper and conclude by mentioning a few topics for further
studies. In Appendix {\bf A}, we present the anisotropic
solutions to general relativity equations. In Appendix {\bf B},
we present the isotropic solutions to the equations in the LQC
-- inspired models. In Appendix {\bf C}, we present solutions
for a case left out in section {\bf 5}.


\vspace{4ex}

\begin{center}

{\bf 2. LQC -- inspired models : Equations of motion}

\end{center}

\vspace{2ex}

In this section, we write down the equations of motion first in
general relativity and then in the Loop Quantum Cosmology (LQC)
-- inspired models. Let the space be $d$ dimensional and
toroidal with $d \ge 3 \;$ and with coordinates $x^i$, $\; i =
1, 2, \cdots, d \;$. Consider a homogeneous and anisotropic
universe whose $(d + 1)$ dimensional line element $d s$ is given
by
\begin{equation}\label{ds}
d s^2 = - \; d t^2 + \sum_i e^{2 \lambda^i} \; (d x^i)^2
\end{equation}
where the scale factors $e^{\lambda^i}$ are functions of $t$
only. Here and in the following, we will explicitly write the
indices to be summed over. The general relativity equations are
given, in the standard notation with $\kappa^2 = 8 \pi G_{d + 1}
\;$, by
\begin{equation}\label{rab}
R_{A B} - \frac{1}{2} \; g_{A B} \; R = \kappa^2 \; T_{A B}
\; \; \; , \; \; \; \;
\sum_A \nabla^A T_{A B} = 0 
\end{equation} 
where $A, B = (0, i)$ and $T_{A B}$ is the energy momentum
tensor. Let $T_{A B}$ be diagonal and be given by $T_{A B} =
diag \; (\rho, \; p_i) \;$ where $\rho$ is the density and $p_i$
is the pressure in the $i^{th}$ direction. Then, after a
straightforward algebra, equations (\ref{rab}) give
\begin{eqnarray}
\sum_{i j} G_{i j} \; \lambda^i_t \; \lambda^j_t
& = & 2 \kappa^2 \; \rho \label{t21} \\
& & \nonumber \\
\lambda^i_{t t} + \Lambda_t \; \lambda^i_t & = & \kappa^2 \;
r^i  \label{t22} \\
& & \nonumber \\
\rho_t + \sum_i (\rho + p_i) \; \lambda^i_t
& = & 0  \label{rhot}
\end{eqnarray}
where the $t-$subscripts denote derivatives with respect to $t
\;$ and 
\begin{eqnarray}
G_{i j} \; = \; 1 - \delta_{i j} & , & G^{i j} \; = \;
\frac{1}{d - 1} - \delta^{i j} \nonumber \\
& & \nonumber \\
\Lambda \; = \; \sum_i \lambda^i & , & 
r^i \; = \; \sum_j G^{i j} \; (\rho - p_j) \; \; . \label{ri}
\end{eqnarray}
Note that $\sum_j G^{i j} G_{j k} = \delta^i_{\; k}$ and $r^i =
p_i + \frac {\rho - \sum_j p_j} {d - 1} \;$. Also define $Y_i$
by
\begin{equation}\label{yi}
Y_i \; = \; \sum_j G_{i j} \; \lambda^j_t
\; = \; \Lambda_t - \lambda^i_t 
\end{equation}
so that, using equation (\ref{t21}), equation (\ref{t22}) for
$\lambda^i_{t t}$ may be written as
\begin{equation}\label{t22y}
\lambda^i_{t t} \; + \;
\sum_j \frac {(\lambda^i_t - \lambda^j_t) \; Y_j} {d - 1}
\; = \; \kappa^2 \;
\left( r^i - \frac {2 \rho} {d - 1} \right) \; \; . 
\end{equation}
Equations (\ref{t21}) and (\ref{t22y}) will resemble closely the
equations (\ref{e1}) and (\ref{e2}) in the LQC -- inspired
model, to be given below.

We now consider the evolution of a $(d + 1)$ dimensional
homogeneous anisotropic universe in the LQC -- inspired
models. These models were constructed in our earlier works by a
natural, straightforward, and empirical generalisation of the
effective equations which describe the quantum evolution of an
anisotropic universe in LQC. The model we consider here is
specified by an arbitrary function $f(x)$ with the only
requirement that $f(x) \to x $ in the limit $x \to 0 \;$. The
general relativity equations follow for $f(x) = x$ and the LQC
effective equations follow for $f(x) = sin \; x$ and $d = 3 \;$.

In the $(3 + 1)$ dimensional Loop Quantum Gravity (LQG)
formalism, the canonical pairs of phase space variables consist
of an $SU (2)$ connection $A^i_a$ and a triad $E^a_i$ of density
weight one where $i, a = 1, 2, 3 \;$. For LQC, in the notation
used here, the triad variable $E^a_i \propto e^{\Lambda -
\lambda^i} \;$ and the connection variable $A^i_a \propto
\hat{c}^i$ which will turn out to be related to $\left(
e^{\lambda^i} \right)_t \;$. Also, let $m^i = \bar{\mu}^i
\hat{c}^i $ where $\bar{\mu}^i \propto e^{- \lambda^i}$ in what
is referred to as the $\bar{\mu}-$scheme. The exact expressions
for $A^i_a$, $\; E^a_i$, and $\bar{\mu}^i$ and their derivations
are somewhat involved and are not needed here. See the review
\cite{status} for a detailed description.

Starting with the LQC variables in $(3 + 1)$ dimensions,
generalising them empirically to $(d + 1)$ dimensions, and after
a long algebra, the equations for the LQC -- inspired models may
be written concisely in terms of the variables $m^i, \; i = 1,
2, \cdots, d \;$. In these models, the conservation equation
(\ref{rhot}) for $\rho_t$ remains the same but equations
(\ref{t21}) and (\ref{t22y}), which is equivalent to
(\ref{t22}), are modified. In terms of the functions $f^i, \;
g_i$, and $X_i$ defined by
\begin{equation}\label{fgx} 
f^i = f(m^i) \; \; , \; \; \;
g_i = \frac{d \; f^i} {d m^i} \; \; , \; \; \;
X_i = g_i \sum_j G_{i j} f^j \; \; ,
\end{equation}
these modified equations in our LQC -- inspired models are given
by
\begin{eqnarray}
\sum_{i j} G_{i j} f^i f^j & = & 2 \; \gamma^2 \lambda_{qm}^2
\kappa^2 \; \rho \; = \; \frac {\rho} {\rho_{qm}} \label{e1} \\
& & \nonumber \\
(m^i)_t \; + \; \sum_j \frac {(m^i - m^j) \; X_j} {(d - 1) \;
\gamma \lambda_{qm}} & = & \gamma \lambda_{qm} \kappa^2 \;
\left( r^i - \frac {2 \rho} {d - 1} \right) \label{e2} \\
& & \nonumber \\ 
\frac {X_i} {\gamma \lambda_{qm}} & = & 
Y_i \; = \; \sum_j G_{i j} \; \lambda^j_t
\nonumber \\
& & \nonumber \\ 
\longleftrightarrow \; \; \; 
\lambda^i_t & = & \frac {\sum_j G^{i j} X_j}
{\gamma \lambda_{qm}} \label{e3}
\end{eqnarray}
where $\rho_{qm} = \frac {1} {2 \; \gamma^2 \lambda_{qm}^2
\kappa^2} \;$, the constant $\gamma$ is analogous to the Barbero
-- Immirzi parameter in LQC, and $\lambda_{qm}$ is a length
parameter which characterises the quantum of the $(d - 1)$
dimensional area : $\lambda_{qm}^{d - 1} \sim \gamma \kappa^2
\;$. Note that, upon using (\ref{e3}) for $\lambda^i_t$ and
equations (\ref{ri}) for $r^i \;$, the conservation equation
(\ref{rhot}) may be written in terms of $X_i$ as

\begin{equation}\label{erhot} 
(\gamma \lambda_{qm}) \; \rho_t \; + \; 2 \rho \;
\frac {\sum_i X_i} {d - 1} \; = \; \sum_i r^i \; X_i \; \; .
\end{equation} 
Equation (\ref{erhot}) also follows upon calculating $\rho_t$
from equation (\ref{e1}) and then using equation (\ref{fgx}) for
$X_i$ and (\ref{e2}) for $(m^i)_t \;$. Equivalently, equation
(\ref{e1}) may be derived as an integral of equations (\ref{e2})
and (\ref{erhot}). Also note that for any linear function $f(x)
= c x + c_0$ where $c$ and $c_0$ are constants, one has
\begin{equation}\label{gic}
f^i \; = \; c m^i + c_0 \; \; , \; \; \;
g_i \; = \; c \; \; , \; \; \;
(\gamma \lambda_{qm}) \; \lambda^i_t \; = \; c f^i \; \; .
\end{equation} 
Equations (\ref{e1}) and (\ref{e2}) then give the general
relativity equations (\ref{t21}) and (\ref{t22y}) with
$\kappa^2$ now replaced by $c^2 \kappa^2 \;$.

We note here that Helling has pointed out in \cite{helling} that
functions of the form $f(x) = \sum_n a_n \; sin \; (b_n x)$
should be admissible within the LQC formalism itself. He further
shows by giving an example that some of these functions with
infinite sums can lead to a more singular evolution than in
general relativity. Also, Bodendorfer et al have constructed a
higher dimensional LQG by generalising Ashtekar variables
\cite{th1, th2, th3}. Upon quantising its homogeneous sector,
one can obtain $(d + 1)$ dimensional LQC where $f(x) = sin \; x$
\cite{zhang, work}. It is likely that functions of the form
$f(x) = \sum_n a_n \; sin \; (b_n x)$ should be admissible here
also. Admitting such functions in $(d + 1)$ dimensional LQC may
provide a firm foundation for the LQC -- inspired models. 


\vspace{4ex}

\centerline{\bf 3. M theory universe}


\vspace{2ex}

One may now study the $(10 + 1)$ dimensional M theory universe
in the LQC -- inspired models by incorporating in equations
(\ref{e1}) -- (\ref{e3}) the density $\rho$ and the pressures
$p_i$ for its constituents.

We consider the M theory universe studied in \cite{m06} --
\cite{k10} which is dominated by constituents that are most
entropic. Such constituents are given by four stacks of M theory
brane--antibranes which intersect according to the Bogomol'nyi
-- Prasad -- Sommerfield (BPS) rules \footnote{According to the
BPS rules, two stacks of 5 branes intersect along three common
spatial directions; two stacks of 2 branes intersect along zero
common spatial directions; a stack of 2 branes intersect a stack
of 5 branes along one common spatial direction; and each stack
of branes is smeared uniformly along the other brane directions.
There can be a wave along common intersection direction. See
\cite{bps, b2ps, b3ps} for more details and for other such M
theory configurations.}  and wrap the seven directions, labelled
$1, 2, \cdots, 7 \;$: namely, two stacks each of $M 2$ and $M 5$
brane--antibranes wrap respectively the directions $1 2$, $\; 3
4$, $\; 1 3 5 6 7$, and $2 4 5 6 7 \;$. Such $N$ stacks of $M 2$
and $M 5$ brane--antibranes intersecting according to the BPS
rules may be described by a total energy momentum tensor $T_{A
B}$ which is made up of $N$ mutually noninteracting and
seperately conserved components. These energy momentum tensors
may be taken to be diagonal. Thus, with $I = 1, 2, \cdots, N
\;$, they may be written as
\begin{equation}\label{tabi}
T_{A B} = \sum_I T_{A B (I)}
\; \; , \; \; \;
\sum_A \nabla^A T_{A B (I)} = 0 
\end{equation}
where $T_{A B} = diag \; (\rho, \; p_i)$ and $T_{A B (I)} = diag
\; (\rho_I, \; p_{i I}) \;$. The total density $\rho$, the total
pressure $p_i$ in the $i^{th}$ direction, and the total $r^i$
are then given by
\begin{equation}\label{totalrho} 
\rho \; = \; \sum_I \rho_I \; \; , \; \; \; 
p_i \; = \;  \sum_I p_{i I} \; \; , \; \; \; 
r^i \; = \;  \sum_I r^i_I 
\end{equation}
where
\begin{equation}\label{riI} 
r^i_I \; = \; \sum_j G^{i j} \; (\rho_I - p_{j I}) \; = \;
p_{i I} + \frac {\rho_I - \sum_j p_{j I}} {d - 1} \; \; .
\end{equation} 
Furthermore, for the line element $d s$ given in equation
(\ref{ds}), the conservation equation (\ref{tabi}) for $T_{A B
(I)}$ leads to
\begin{equation}\label{rhoIt} 
(\rho_I)_t + \sum_i (\rho_I + p_{i I}) \; \lambda^i_t \; = \; 0
\; \; . 
\end{equation}
In the LQC -- inspired models, using equations (\ref{e3}) for
$\lambda^i_t$ and (\ref{riI}) for $r^i_I \;$, the conservation
equation (\ref{rhoIt}) may be written in terms of $X_i$ as
\begin{equation}\label{erhoIt} 
(\gamma \lambda_{qm}) \; (\rho_I)_t \; + \; 2 \rho_I \; \frac
{\sum_i X_i} {d - 1} \; = \; \sum_i r^i_I \; X_i \; \; . 
\end{equation}

To proceed further, one needs equations of state which determine
the pressures $p_{i I}$ in terms of $\rho_I \;$. For $N$ stacks
of $M2$ and $M5$ brane--antibranes intersecting according to the
BPS rules, the U--duality symmetries of M theory may be shown
\cite{k207, k08, k10} to require that the density $\rho_{(I)}$ of
the $I^{th}$ stack and its pressures $p_{\parallel (I)}$ and
$p_{\perp (I)}$ along the parallel and transverse directions
must be related as follows :
\begin{equation}\label{pp}
p_{\parallel (I)} = - \rho_{(I)} + 2 \; p_{\perp (I)}
\; \; \; \longleftrightarrow \; \; \;
(\rho - p_\parallel)_{(I)} = 2 \; (\rho - p_\perp)_{(I)} \; \; .
\end{equation}
Specifying $p_{\perp (I)}$ as a function of $\rho_{(I)}$ will
determine the equations of state for $p_{\parallel (I)}$ and
thereby for all the pressures $p_{i (I)} \;$. The U--duality
symmetries further require this function to be the same for all
$I \;$. Hence, specifying a single function $p_\perp (\rho)$
determines all $p_{i I}$ in terms of $\rho_I$ where $i = 1, 2,
\cdots, 10$ and $I = 1, 2, \cdots, N \;$. \footnote{ In a
certain approximation, Chowdhury and Mathur have derived from
first principles the energy momentum tensors for the
intersecting branes \cite{m06, m08}. The pressures, thus
derived, satisfy the U--duality relation (\ref{pp}) and follow
from the present expressions as a special case when $p_\perp = 0
\;$.}

Consider now the most entropic constituents mentioned earlier
which are given by two stacks each of $M 2$ and $M 5$
brane--antibranes. $N = 4$ for this configuration and, for
simplicity, we refer to it as $(2, \; 2', \; 5, \; 5')$ branes.
Using equation (\ref{pp}), we now write the pressures in the
$i^{th}$ directions for the $(2, \; 2', \; 5, \; 5')$ branes in
an obvious notation as follows:
\begin{eqnarray} 
\{ (\rho - p_i)_{(2)} \} \; :
& (2, \; 2, \; 1, \; 1, \; 1, \; 1, \; 1, \; 1, \; 1, \; 1) \; 
(\rho - p_\perp)_{(2)} & \nonumber \\
& & \nonumber \\
\{ (\rho - p_i)_{(2')} \} \; :
& (1, \; 1, \; 2, \; 2, \; 1, \; 1, \; 1, \; 1, \; 1, \; 1) \; 
(\rho - p_\perp)_{(2')} & \nonumber \\
& & \nonumber \\
\{ (\rho - p_i)_{(5)} \} \; :
& (2, \; 1, \; 2, \; 1, \; 2, \; 2, \; 2, \; 1, \; 1, \; 1) \; 
(\rho - p_\perp)_{(5)} & \nonumber \\
& & \nonumber \\
\{ (\rho - p_i)_{(5')} \} \; :
& (1, \; 2, \; 1, \; 2, \; 2, \; 2, \; 2, \; 1, \; 1, \; 1) \;
(\rho - p_\perp)_{(5')} & . \label{pi22'55'}
\end{eqnarray}
The corresponding $r^i_{(*)} = \sum_j G^{i j} \; (\rho -
p_j)_{(*)}$ where $* = 2, \; 2', \; 5, \; 5'$ are given, after a
little algebra, by
\begin{eqnarray}
\{ r^i_{(2)} \} \; : & (- 2, \; - 2, \; 1, \; 1, \; 1, \; 1, \;
1, \; 1, \; 1, \; 1) \;
\frac {(\rho - p_\perp)_{(2)}} {3} & \nonumber \\
& & \nonumber \\
\{ r^i_{(2')} \} \; : & (1, \; 1, \; - 2, \; - 2, \; 1, \; 1, \;
1, \; 1, \; 1, \; 1) \;
\frac {(\rho - p_\perp)_{(2')}} {3} & \nonumber \\
& & \nonumber \\
\{ r^i_{(5)} \} \; : & (- 1, \; 2, \; - 1, \; 2, \; - 1, \; - 1,
\; - 1, \; 2, \; 2, \; 2) \;
\frac {(\rho - p_\perp)_{(5)}} {3} & \nonumber \\
& & \nonumber \\
\{ r^i_{(5')} \} \; : & (2, \; - 1, \; 2, \; - 1, \; - 1, \; -
1, \; - 1, \; 2, \; 2, \; 2) \;
\frac {(\rho - p_\perp)_{(5')}} {3} & . \label{ri22'55'}
\end{eqnarray}
Thus, given an equation of state function $p_\perp (\rho) \;$,
equations (\ref{e1}) -- (\ref{e3}), (\ref{totalrho}) --
(\ref{erhoIt}), and (\ref{ri22'55'}) will describe the
cosmological evolution of a $(10 + 1)$ dimensional M theory
universe in our LQC -- inspired models.

Note that if the densities $\rho_{(*)}$ are the same for all $*
= 2, \; 2', \; 5, \; 5'$ then so will be the pressures $p_{\perp
(*)}$ and, hence, $(\rho - p_\perp)_{(*)} \;$. Consequently, it
follows from the above expressions for $r^i_{(*)}$ that the
total $r^i = r^i_{(2)} + r^i_{(2')} + r^i_{(5)} + r^i_{(5')} =
0$ for $i = 1, 2, \cdots, 7 \;$. The ten dimensional space will
then become effectively three dimensional in the limit
$e^\Lambda \to \infty \;$ : the seven directions, labelled $1,
2, \cdots, 7 \;$, will neither expand nor contract and will
reach constant sizes; and, the remaining three directions will
continue to expand. In this paper, we assume that the densities
$\rho_{(*)}$ are the same for all $*$ and that the equation of
state is linear. \footnote{Even if the densities $\rho_{(*)}$
are unequal initially, the dynamics of the general relativity
equations (\ref{t22}) resulting from the $r^i_{(*)}$ given in
equations (\ref{ri22'55'}) is such that these densities become
equal in the limit $e^\Lambda \to \infty \;$ \cite{k207, k08,
k10}. Such an M theory universe may therefore provide a detailed
realisation of the maximum entropic principle that we had
proposed in \cite{k206} to determine the number (3 + 1) of large
spacetime dimensions.} Thus, we write
\begin{equation}\label{eos}
\rho_{(*)} \; = \; \frac {\rho} {4} \; \; , \; \; \; 
p_{\perp (*)} \; = \; (1 - u) \; \rho_{(*)}
\end{equation}
for $* = 2, \; 2', \; 5, \; 5'$ where $\rho$ is the total
density and $u < 2$ is a constant. It then follows from
equations (\ref{pi22'55'}) and (\ref{ri22'55'}) that the total
$p_i = \sum_I p_{i I}$ and $r^i = \sum_I r^i_I$ are given by
\begin{eqnarray}
\{ \rho - p_i \} & : & (6, \; 6, \; 6, \; 6, \; 6, \; 6, \; 6,
\; 4, \; 4, \; 4) \; \; \frac {u \; \rho} {4}
\label{pibi} \\ 
& & \nonumber \\
\{ r^i \} & : & (0, \; 0, \; 0, \; 0, \; 0, \; 0, \; 0, \;
6, \; 6, \; 6) \; \; \frac {u \; \rho} {12} \; \; . \label{ribi}
\end{eqnarray}


\vspace{4ex}

\centerline{\bf 4. General evolution and a bi--anisotropic
universe}

\vspace{2ex}

Consider now the general evolution resulting from equations
(\ref{e1}) -- (\ref{e3}) and (\ref{totalrho}) -- (\ref{erhoIt}).
Equation (\ref{e1}) may be derived as an integral of the
remaining equations. Hence, if it is satisfied at an initial
time $t_0$ then it is satisfied for all $t \;$.

The equations of state, which may be derived from the underlying
physics or may be assumed, will give the pressures $p_{i I}$ and
the quantities $r^i_I$ in terms of $\rho_I \;$. Then equations
(\ref{e2}), (\ref{e3}), and (\ref{erhoIt}) give the first time
derivatives $m^i_t \;$, $\; \lambda^i_t \;$, and $(\rho_I)_t \;$
as polynomials in terms of $(\rho_I, \; m^i, \; f^i)$ and $g_i =
\frac {d \; f^i} {d m^i}$ where $f^i = f(m^i) \;$.
Differentiating these expressions repeatedly will then give all
the higher time derivatives of $(m^i, \; \lambda^i, \;\rho_I)$
as polynomials in terms of $(\rho_I, \; m^i, \; f^i)$ and the
higher derivatives of $f^i$ with respect to $m^i \;$. Therefore,
it follows that if the function $f(x)$ and all its derivatives
are finite then all the time derivatives of $\lambda^i$ will
also remain finite and thus the evolution will be non singular.
See \cite{k17} for a variety of such evolutions. Also, note that
the function $f(x) = x$ is not finite although all its
derivatives are, and it leads to the big bang singularities of
general relativity.

Consider obtaining solutions numerically for $(m^i, \;
\lambda^i, \; \rho_I) \;$. Let the equations of state be given
and let the initial values of $(m^i, \; \lambda^i, \; \rho_I)$
at $t_0$ satisfying equation (\ref{e1}) be also given. Then, in
principle, $m^i (t)$, $\; \lambda^i (t)$, and $\rho_I (t) \;$
follow from equations (\ref{e2}), (\ref{e3}), and (\ref{erhoIt})
: The values of $m^i (t_0)$ determine the values of $(f^i, \;
g_i, \; X_i)$ at $t_0 \;$; equation (\ref{e3}) then determines
$\lambda^i_t$ at $t_0 \;$; and equations (\ref{e2}) and
(\ref{erhoIt}), together with the equations of state, then
determine $m^i_t$ and $(\rho_I)_t$ at $t_0 \;$. These will then
determine the values of $(m^i, \; \lambda^i, \; \rho_I)$ at $t_0
\pm \delta t \;$. Repeating this procedure will give $(m^i, \;
\lambda^i, \; \rho_I)$ for all $t \;$. Thus, it is always
possible to obtain solutions numerically.

However, solving equations (\ref{e1}) -- (\ref{e3}) and
(\ref{totalrho}) -- (\ref{erhoIt}) analytically and obtaining
$m^i (t)$, $\; \lambda^i (t)$, and $\rho_I (t)$ explicitly is
not always possible. We are able to obtain explicit solutions
only in a few simple cases when $N = 1 \;$ and when the equation
of state is linear : in the anisotropic case with $f(x) = c x +
c_0$ which gives general relativity, see Appendix {\bf A}; and,
in the isotropic case with $f(x) = c x + c_0$ or $f(x) = sin \;
x \;$, see \cite{k16, k17} and Appendix {\bf B}.

Hence, in order to obtain explicit solutions which may provide
insights into non singular evolution of an M theory universe, we
now simplify our set up : Instead of considering a general
anisotropic universe, we consider a bi--anisotropic universe
where the space is $d = (\tilde{n} + n)$ dimensional, and where
the quantities, such as $m^i, \; f^i, \; \lambda^i, \; p_i, \;
r^i$, corresponding to the $\tilde{n}$ and the $n$ dimensional
spaces are seperately isotropic. Thus, we write
\begin{eqnarray}
& & \left( m^i, \; f^i, \; g_i, \; X_i, \; \lambda^i, \; p_i, \;
r^i \right) \nonumber \\
& & \nonumber \\
& = & \left( \tilde{m}, \; \tilde{f}, \; \tilde{g}, \;
\tilde{X}, \; \tilde{\lambda}, \; \tilde{p}, \; \tilde{r}
\right) \; \; \; for \; \; \; i = 1, 2, \cdots, \tilde{n}
\nonumber \\
& & \nonumber \\
& = & \left( m, \; f, \; g, \; X, \; \lambda, \; p, \; r \right)
\; \; \; for \; \; \; i = \tilde{n} + 1, \cdots, \tilde{n} + n
\; \; . \label{lqtll}
\end{eqnarray}
Then the line element $d s$ in equation (\ref{ds}) is given by
\begin{equation}\label{dstll}
d s^2 = - \; d t^2 + e^{2 \tilde{\lambda}} \;
\sum_{i = 1}^{\tilde{n}} (d x^i)^2 + e^{2 \lambda} \;
\sum_{i = \tilde{n} + 1}^{\tilde{n} + n} \; (d x^i)^2 \; \; ,
\end{equation}
we have $\Lambda = \tilde{n} \tilde{\lambda} + n \lambda \;$,
and equations (\ref{e1}) -- (\ref{e3}) become
\begin{eqnarray}
\left( n f + \tilde{n} \tilde{f} \right)^2 - \left(
n f^2 + \tilde{n} \tilde{f}^2 \right) & = &
2 \; \gamma^2 \lambda_{qm}^2 \kappa^2 \; \rho
\; = \; \frac {\rho} {\rho_{qm}} \label{te1} \\
& & \nonumber \\
\tilde{m}_t \; + \; \frac {(\tilde{m} - m) \; n X} {(d - 1) \;
\gamma \lambda_{qm}} & = & \gamma \lambda_{qm} \kappa^2 \;
\left( \tilde{r} - \frac {2 \rho} {d - 1} \right) \nonumber \\
& & \nonumber \\
m_t \; + \; \frac {(m - \tilde{m}) \; \tilde{n} \tilde{X}}
{(d - 1) \; \gamma \lambda_{qm}} & = & \gamma \lambda_{qm}
\kappa^2 \; \left( r - \frac {2 \rho} {d - 1} \right)
\label{te2} \\
& & \nonumber \\
\frac {\tilde{X}} {\gamma \lambda_{qm}} \; = \; \Lambda_t
- \tilde{\lambda}_t & , & \frac {X} {\gamma \lambda_{qm}}
\; = \; \Lambda_t - \lambda_t \nonumber \\
& & \nonumber \\
\longleftrightarrow \; \; \; \tilde{\lambda}_t \; = \;
\frac {n X - (n - 1) \tilde{X}} {(d- 1) \;
(\gamma \lambda_{qm})} & , & \lambda_t \; = \; \frac
{\tilde{n} \tilde{X} - (\tilde{n} - 1) X}
{(d - 1) \; (\gamma \lambda_{qm})} \label{te3}
\end{eqnarray}
where
\begin{eqnarray}
\tilde{X} = \tilde{g} \; \left(
n f + (\tilde{n} - 1) \tilde{f} \right) & , & X = g \;
\left( (n - 1) f + \tilde{n} \tilde{f} \right) \nonumber \\
& & \nonumber \\
\tilde{r} = \frac {\rho - n p + (n - 1) \tilde{p}} {d - 1}
& , & r = \frac {\rho - \tilde{n} \tilde{p} + (\tilde{n} - 1) p}
{d - 1} \; \; . \label{tXX}
\end{eqnarray}

Let the equations of state be
linear and be given by $\tilde{p} = (1 - \tilde{u}) \; \rho$ and
$p = (1 - u) \; \rho \;$. Then, writing $\tilde{r} = \tilde{v}
\rho$ and $r = v \rho \;$, one has
\begin{equation}\label{tvv}
\tilde{v} \; = \; \frac {n u - (n - 1) \tilde{u}} {d - 1}
\; \; \; , \; \; \; \;
v \; = \; \frac {\tilde{n} \tilde{u} - (\tilde{n} - 1) u}
{d - 1} \; \; .
\end{equation}
For the $(\tilde{n} + n)$ dimensional space to become
effectively $n$ dimensional in the limit $e^\Lambda \to \infty
\;$, it is necessary that $\tilde{v} = 0$ which then gives
\begin{equation}\label{tv0v}
\tilde{u} \; = \; \frac {n \; u} {n - 1}
\; \; \; , \; \; \; \;
v \; = \; \frac {u} {n - 1} \; \; .
\end{equation}
Also, for the linear equations of state, the conservation
equation (\ref{rhot}) gives
\begin{equation}\label{rhotll}
\rho \; = \; \rho_0 \; e^{ (\tilde{u} - 2) \;
\tilde{n} (\tilde{\lambda} - \tilde{\lambda}_0)
\; + \; (u - 2) \; n (\lambda - \lambda_0) } \; \; .
\end{equation} 

For an M theory universe, $d = \tilde{n} + n = 10 \;$. The above
equations for the bi--anisotropic universe are consistent with
and become applicable to M theory universe dominated by $(2, \;
2', \; 5, \; 5')$ branes if $\tilde{n} = 7 \;$, $ \; n = 3 \;$,
and the densities $\rho_{(*)}$ are the same for all $* = 2, \;
2', \; 5, \; 5' \;$. Therefore, we take $\rho_{(*)}$ and the
equation of state to be given by equations (\ref{eos}).  Hence
$p = (1 - u) \rho \;$. Then, with $\tilde{p} = (1 - \tilde{u})
\rho \;$, $\; \tilde{r} = \tilde{v} \rho \;$, and $r = v \rho
\;$, one has $\tilde{u} = \frac {3 u} {2} \;$, $\; \tilde{v} = 0
\;$, and $v = \frac {u} {2} \;$, see equations (\ref{pibi}),
(\ref{ribi}), and (\ref{tv0v}).


\vspace{4ex}

\begin{center}

{\bf A convenient choice for $f(x)$ }

\end{center}

\vspace{2ex}

In the LQC -- inspired models, it follows from equations
(\ref{e1}) -- (\ref{e3}) and (\ref{totalrho}) -- (\ref{erhoIt})
that the cosmological evolution will be non singular if the
function $f(x)$ and all its derivatives are finite. We do not
know the fundamental origin, if any, of such a class of
functions. Nevertheless, by modelling the non singular evolution
of an universe in several ways by several choices of $f(x) \;$,
one may gain new insights into the Planckian regime of the
evolution.

One question that may be asked in the present set up is the
following. In an M theory universe where the constituent
pressures are given by equation (\ref{pi22'55'}), the seven
spatial directions wrapped by branes reach constant sizes and
the remaining three continue to expand in the limit $e^\Lambda
\to \infty \;$. As we found in \cite{k08, k10} using general
relativity equations, these constant sizes are generically of
${\cal O} (l_{11})$ where $l_{11}$ is the eleven dimensional
Planck length. They may be made arbitrarily large, for example
${\cal O} (10^{15} \; l_{11})$ which may be of phenomenological
interest \cite{add, cq}, but it requires a similary large fine
tuning to about $15$ decimal places near the Planckian
regime. One may now ask in an LQC -- inspired model for an M
theory universe whether it is possible to obtain a large
internal space with no fine tuning. 

Such a question may be addressed in the LQC -- inspired models
because now the evolution can be made non singular by choosing
the function $f(x)$ appropriately.  Naturally, one may also hope
to achieve a large internal space but with no fine tuning by
choosing a suitable class of such functions. With this question
in mind, we consider functions which may cause the universe to
be in the Planckian regime for a long time and study whether a
long stay in the Planckian regime will result in a large
internal space.

Accordingly, we consider a class of functions which are odd
under $x \to - x \;$, have a period $4 m_* \;$, are labelled by
an integer $\nu \ge 1 \;$, and are given in the interval $0 \le
x \le 2 m_*$ by
\begin{equation}\label{fnu}
f(x ;  \nu) \; = \; A \; 
\left( 1 - \left( 1 - \frac {x} {m_*} \right)^{2 \nu} \right)
\end{equation}
where $m_* = 2 \nu A$ so that $f(x ; \nu) \to x $ in the limit
$x \to 0 \;$. One may set $A = 1$ with no loss of generality but
it is convenient not to do so. Note that $f(x ; \nu) = 0$ at $x
= 0$ and $2 m_* \;$, that $f(x; \nu) = f_{max} = A$ at $x = m_*
\;$, and that the integer $\nu$ controls the flatness of $f(x;
\nu)$ near its maximum. Also note that when one or more $f^i$s
are of ${\cal O} (1) \;$ and near $f_{max} \;$, equations
(\ref{e1}) and (\ref{e3}) imply that, generically, the values of
$\rho$ and $\lambda^i_t$ are Planckian. Thus, larger values of
$\nu$ will make the function flatter near the maximum and,
hence, may cause the universe to be in the Planckian regime for
a longer time.

Now, in order to obtain explicit solutions, we make a piece-wise
linear approximation to this function as follows. Let $f(- x) =
- f(x) \;$, let $f(x + 4 m_*) = f(x) \;$, and let $f(x)$ be
given in the interval $0 \le x \le 2 m_*$ by
\begin{eqnarray}
f(x) \; = \; x \; \; \; & for & \; \; 0 \le x \le A
\nonumber \\
& & \nonumber \\
\; = \; A \; \; \; & for & \; \; A \le x \le A + 2 \Delta
\nonumber \\
& & \nonumber \\
\; = \; (2 m_* - x) \; \; \; & for & \; \; 
A + 2 \Delta \le x \le 2 m_* \label{flin}
\end{eqnarray}
where $m_* = A + \Delta \;$. The parameter $\Delta$ controls the
width of the flat part of $f(x)$ and, in that sense, is a proxy
for $\nu \;$. Note that the functions given in equations
(\ref{fnu}) and (\ref{flin}) have discontinuities in their
derivatives which are but artefacts of our modelling. We will
ignore such discontinuities because they may all be smoothened
out as much as required. Then, since the function remains finite
and all its derivatives may be smoothened to finite values, the
resulting evolution will be non singular.


\vspace{4ex}

\begin{center}

{\bf 5. Solutions for a bi--anisotropic universe}

\end{center}

\vspace{2ex}

Consider the solutions to equations (\ref{e1}) -- (\ref{e3})
when $f(x)$ is the simplified, piece-wise linear function given
in equation (\ref{flin}). Isotropic solutions are
straightforward to obtain and they are given in Appendix {\bf
B}. Consider the solutions for a bi-anisotropic universe where
$d = \tilde{n} + n$ and the quantities corresponding to the
$\tilde{n}$ and the $n$ dimensional spaces are seperately
isotropic as given in equation (\ref{lqtll}).

Equations (\ref{te1}) -- (\ref{te3}) describe the evolution of
such an universe. Let the equations of state be given by
$\tilde{p} = (1 - \tilde{u}) \rho$ and $p = (1 - u) \rho \;$.
Then $\tilde{r} = \tilde{v} \rho$ and $r = v \rho$ where
$\tilde{v}$ and $v$ are given by equations (\ref{tvv}), and
equation (\ref{rhotll}) gives $\rho$ in terms of
$\tilde{\lambda}$ and $\lambda \;$. If $\tilde{m}$ lies in the
interval $(0, \; A)$ and $m$ in $(A + 2 \Delta, \; 2 m_*)$ or
vice versa, then we cannot solve equations (\ref{te1}) and
(\ref{te2}) analytically. Hence we assume that $\Delta \gg A$ so
that, generically, this possibility will not arise.

When $\tilde{m}$ and $m$ both lie in the interval $(0, \; A)$,
the evolution will be as in general relativity for which the
solutions are given in Appendix {\bf A}. Let $t_0$ be an initial
time and let the initial values $\tilde{m}_0$ and $m_0$ both lie
in the interval $(0, \; A) \;$. Then the initial values
$\tilde{\lambda}_{t 0}$ and $\lambda_{t 0}$ are both positive,
see equations (\ref{gic}). Hence, going forward in time,
$\tilde{m} \propto \tilde{\lambda}_t$ and $m \propto \lambda_t$
will decrease monotonically for $t > t_0$ and will vanish in the
limit $t \to \infty \;$.

Going back in time, $\tilde{m}$ and $m$ will increase
monotonically for $t < t_0 \;$. They will enter the interval
$(A, \; A + 2 \Delta)$ one after the other, evolve further, and
exit from it into the interval $(A + 2 \Delta, \; 2 m_*)
\;$. Let these entries and exits occur at times $(t_{\tilde{1}},
\; t_1, \; t_{\tilde{2}}, \; t_2) \;$. Taking $t_0 >
t_{\tilde{1}} > t_1 > t_{\tilde{2}} > t_2$ for the sake of
definiteness, we denote the monotonically increasing values of
$\tilde{m}$ and $m$ at these times by
\begin{equation}\label{mtimes}
(\tilde{m}_0, \; \tilde{m}_{\tilde{1}}, \; \tilde{m}_1, \;
\tilde{m}_{\tilde{2}}, \; \tilde{m}_2) \; \; \; , \; \; \;
(m_0, \; m_{\tilde{1}}, \; m_1, \; m_{\tilde{2}}, \; m_2)
\end{equation}
where
\begin{eqnarray}
0 \; < \; \tilde{m}_0 \; < \; A & , & 0 \; < \; m_0 \; < \; A
\nonumber \\
& & \nonumber \\
\tilde{m}_{\tilde{1}} \; = \; A & , & m_{\tilde{1}} \; < \; A
\nonumber \\
& & \nonumber \\
A \; < \; \tilde{m}_1 \; < \; A + 2 \Delta & , & m_1 \; = \; A
\nonumber \\
& & \nonumber \\
\tilde{m}_{\tilde{2}} \; = \; A + 2 \Delta & , &
A \; < \; m_{\tilde{2}} \; < \; A + 2 \Delta \nonumber \\
& & \nonumber \\
\tilde{m}_2 \; > \; A + 2 \Delta & , & m_2 \; = \; A + 2 \Delta
\; \; . \label{AA}
\end{eqnarray}
Also, let the values of $\tilde{\lambda}$ and $\lambda$ at the
times $(t_0, \; t_{\tilde{1}}, \; t_1, \; t_{\tilde{2}}, \;
t_2)$ be denoted by
\begin{equation}\label{ltimes}
(\tilde{\lambda}_0, \; \tilde{\lambda}_{\tilde{1}}, \;
\tilde{\lambda}_1, \; \tilde{\lambda}_{\tilde{2}}, \;
\tilde{\lambda}_2) \; \; \; , \; \; \;
(\lambda_0, \; \lambda_{\tilde{1}}, \; \lambda_1, \;
\lambda_{\tilde{2}}, \; \lambda_2) \; \; .
\end{equation}
In expressions (\ref{AA}), the equalities define the times
$(t_{\tilde{1}}, \; t_1, \; t_{\tilde{2}}, \; t_2)$ and the
inequalities mean that, as one goes back in time from $t_0 \;$,
the field $\tilde{m}$ first enters the interval $(A, \; A + 2
\Delta)$ at $ \tilde{t}_1 \;$, then $m$ enters it at $t_1 \;$,
then $\tilde{m}$ first exits from it into the interval $(A + 2
\Delta, \; 2 m_*)$ at $ \tilde{t}_2 \;$, and then $m$ does the
same at $t_2 \;$. We now analyse the solutions as $t$ varies
from $\infty$ to $t_0$ to $t_{\tilde{1}}$ to $t_1$ to
$t_{\tilde{2}}$ to $t_2$ to $- \infty \;$.


\vspace{2ex}

\centerline{\bf $\mathbf{ t > t_{\tilde{1}}}$}

\vspace{2ex}

The fields $\tilde{m}$ and $m$ both lie in the interval $(0, \;
A)$ for $t > t_{\tilde{1}} \;$ and, hence, $f(x) = x \;$.
Therefore, their evolution during these times will be as in
general relativity. The initial values of the fields given at
$t_0 > t_{\tilde{1}}$ and the general relativity solutions given
in Appendix {\bf A} will determine all the fields for $t >
t_{\tilde{1}} \;$. In particular, the values
$\tilde{\lambda}_{\tilde{1}}$ and $\lambda_{\tilde{1}}$ in
expressions (\ref{ltimes}) will follow from these solutions.


\vspace{2ex}

\centerline{\bf $\mathbf{t_{\tilde{1}} > t > t_1}$}

\vspace{2ex}

During $t_{\tilde{1}} > t > t_1 \;$, the field $\tilde{m}$ lies
in the interval $(A, \; A + 2 \Delta)$ and varies from
$\tilde{m}_{\tilde{1}} = A$ to $\tilde{m}_1 > A$ wheres $m$ lies
in the interval $(0, \; A)$ and varies from $m_{\tilde{1}} < A$
to $m_1 = A \;$. Therefore, during this evolution, $\tilde{f} =
A$, $\; \tilde{g} = \tilde{X} = 0$, $ \; f = m$, $\; g = 1$, and
$X = (n - 1) f + \tilde{n} A \;$. Define $y, \; z$, and $a$ by
\begin{equation}\label{yz1} 
y = (n - 1) \; f + \tilde{n} \; A \; \; , \; \; \;
z = (\tilde{m} - m) \; \; , \; \; \; 
a = \sqrt{\frac {\tilde{n} \; (d - 1)} {n}} \; A  \; \; .
\end{equation}
Then, after a straightforward algebra, it follows from equations
(\ref{te1}) and (\ref{te2}) that
\begin{eqnarray}
\frac {\rho} {\rho_{qm}} & = &
\frac {n} {n - 1} \left( y^2 - a^2 \right) \label{rhoy2} \\
& & \nonumber \\
y_t & = & - \; c_y \; (y^2 - a^2) \label{yt} \\
& & \nonumber \\
z_t + b \; y \; z & = & - \; c_z \; (y^2 - a^2)  \label{zt}
\end{eqnarray}
where
\[
c_y = \frac {n \; \left( \frac {2} {d - 1} - v \right)}
{2 \; \gamma \lambda_{qm}} \; \; , \; \; \;
c_z = \frac {n \; (v - \tilde{v})}
{2 \; (n - 1) \; \gamma \lambda_{qm}} \; \; , \; \; \;
b = \frac {n} {(d - 1) \; \gamma \lambda_{qm}} \; \; .
\]  
Since $\tilde{X} = 0$, it follows that $\Lambda_t -
\tilde{\lambda}_t = 0$ and hence, from equations (\ref{tvv}),
(\ref{rhotll}), and (\ref{rhoy2}), that
\begin{eqnarray}
(\lambda - \lambda_{\tilde{1}}) & = & - \;
\left( \frac {\tilde{n} - 1} {n} \right)\;
(\tilde{\lambda} - \tilde{\lambda}_{\tilde{1}}) \nonumber \\
& & \nonumber \\
e^{(2 \; - \; (d - 1) \; v) \; (\tilde{\lambda}_{\tilde{1}}
- \tilde{\lambda})} & = & \frac {\rho} {\rho_{\tilde{1}}}
\; = \; \frac {y^2 - a^2} {y^2_{\tilde{1}} - a^2}
\; \; . \label{tx0} 
\end{eqnarray}
Defining $t_\infty$ by $y (t_\infty) = \infty \;$, the solution
$y (t)$ for the equation (\ref{yt}) may be written as
\begin{equation}\label{ysoln}
\left( \frac {y - a} {y + a} \right) \; e^{2 a c_y \; t} \; = \;
\left( \frac {y_{\tilde{1}} - a} {y_{\tilde{1}} + a} \right) \;
e^{2 a c_y \; t_{\tilde{1}}}
\; = \; \left( \frac {y_1 - a} {y_1 + a} \right) \;
e^{2 a c_y \; t_1} \; = \; e^{2 a c_y \; t_\infty} 
\end{equation}
where $y \ge a > 0$, $\; y_1 = (d - 1) A$, and the last two
equalities give $t_1$ and $t_\infty$ in terms of $A$ and the
initial values $t_{\tilde{1}}$ and $y_{\tilde{1}} \;$. Thus, if
$c_y$ is positive then $y_t < 0 \;$, $\; t_\infty <
t_{\tilde{1}} \;$, and $y$ increases monotonically from $a$ to
$y_{\tilde{1}}$ to $\infty$ as $t$ decreases from $\infty$ to
$t_{\tilde{1}}$ to $t_1$ to $t_\infty \;$. Defining $s = \frac
{b} {2 c_y} \;$ and solving for $z$ in terms of $y$, it is easy
to see that the solution for $z(y)$ is given by
\begin{equation}\label{zsoln}
z \; = \; (y^2 - a^2)^s \; \left( \frac {z_{\tilde{1}}}
{(y^2_{\tilde{1}} - a^2)^s} \; + \; \frac {c_z} {c_y} \;
\int_{y_{\tilde{1}}}^y \; \frac {d y} {(y^2 - a^2)^s} \right)
\; \; .
\end{equation}


\vspace{2ex}

\centerline{\bf $\mathbf{ t_1 > t > t_{\tilde{2}} }$}

\vspace{2ex}

The fields $\tilde{m}$ and $m$ both lie in the interval $(A, \;
A + 2 \Delta)$ when $t$ decreases from $t_1$ to $t_{\tilde{2}}
\;$. It then follows that
\[
\tilde{f} = f = A \; \; \; ,  \; \; \; \; 
\tilde{g} = g = \tilde{X} = X = 0 \; \; . 
\]
Equations (\ref{te3}) then give
\begin{equation}\label{txx0}
\tilde{\lambda}_t \; = \;  \lambda_t \; = \; 0 
\; \; \; \Longrightarrow \; \; \; 
\tilde{\lambda}_{\tilde{2}} \; = \; \tilde{\lambda}_1
\; \; , \; \; \;
\lambda_{\tilde{2}} \; = \; \lambda_1 \; \; . 
\end{equation}
Equations (\ref{te1}) and (\ref{te2}) give
\begin{eqnarray}
\frac {\rho} {\rho_{qm}} & = &
d \; (d - 1) \; A^2 \label{0te1} \\
& & \nonumber \\
\tilde{m} - \tilde{m}_1 & = & c_{\tilde{m}} \; (t_1 - t)
\nonumber \\ 
& & \nonumber \\
m - m_1 & = & c_m \; (t_1 - t) \label{0te2} 
\end{eqnarray}
where $c_{\tilde{m}} = \frac {d (d - 1) A^2} {2 \gamma
\lambda_{qm}} \left( \frac {2} {d - 1} - \tilde{v} \right) \;$,
$\; c_m = \frac {d (d - 1) A^2} {2 \gamma \lambda_{qm}} \left(
\frac {2} {d - 1} - v \right) \;$, $\; \tilde{m}_1 > A \;$, and
$m_1 = A \;$. We will assume that $\tilde{v}$ and $v$ are both
$< \frac {2} {d - 1} \;$, hence $c_{\tilde{m}}$ and $c_m$ are
both positive. There is no loss of generality here since this is
Planckian regime and the constituents with lowest $\tilde{u}$
and $u$ will dominate. Also, for an M theory universe,
$\tilde{v} = 0$ which is clearly $< \frac {2} {d - 1} \;$.
Evolving as in equation (\ref{0te2}), $\tilde{m}$ and $m$ will
reach the value $(A + 2 \Delta)$ respectively at $t_{\tilde{2}}$
and $t_2$ given by
\begin{equation}\label{0'te3}
t_1 - t_{\tilde{2}} \; = \; \frac {A + 2 \Delta - \tilde{m}_1}
{c_{\tilde{m}}} \; \; , \; \; \;
t_1 - t_2 \; = \; \frac {2 \Delta} {c_m} \; \; .
\end{equation}
If $\tilde{v} = v$ then $c_{\tilde{m}} = c_m$ and, since
$\tilde{m}_1 > A$, it follows that $t_{\tilde{2}} > t_2 \;$. If
$\Delta \;$ is large so that $2 \Delta \gg \tilde{m}_1 - A$ then
$A + 2 \Delta - \tilde{m}_1 \simeq 2 \Delta \;$ and
\begin{equation}\label{0te4}
\frac {t_1 - t_{\tilde{2}}} {t_1 - t_2} \; \simeq \;
\frac {c_m} {c_{\tilde{m}}} \; = \;
\frac {2 - (d - 1) \; v} {2 - (d - 1) \; \tilde{v}} \; \; .
\end{equation}

Hence, it follows that $t_{\tilde{2}} > t_2$ if $v > \tilde{v}$
and that $t_2 > t_{\tilde{2}}$ if $\tilde{v} > v \;$. Since we
have assumed that $t_{\tilde{2}} > t_2$, we must have $v \ge
\tilde{v} \;$.


\vspace{2ex}

\centerline{\bf $\mathbf{t_{\tilde{2}} > t > t_2}$}

\vspace{2ex}

During $t_{\tilde{2}} > t > t_2 \;$, the field $m$ lies in the
interval $(A, \; A + 2 \Delta)$ and varies from $m_{\tilde{2}} <
A + 2 \Delta$ to $m_2 = A + 2 \Delta$ whereas $\tilde{m}$ lies
in the interval $(A + 2 \Delta, \; 2 m_*)$ and varies from
$m_{\tilde{2}} = A + 2 \Delta$ to $\tilde{m}_2 > A + 2 \Delta
\;$. The corresponding solutions are similar to the ones when
$t_{\tilde{1}} > t > t_1 \;$. They are given in Appendix {\bf
C}.


\vspace{2ex}

\centerline{\bf $\mathbf{t_2 > t}$}

\vspace{2ex}

The fields $\tilde{m}$ and $m$ both lie in the interval $(A + 2
\Delta, \; 2 m_*)$ for $t < t_2 \;$ and, hence, $f(x) = 2 m_* -
x \;$. Therefore, their evolution during these times will be as
in general relativity. The values of the fields at $t_2 \;$,
equation (\ref{gic}), and the general relativity solutions given
in Appendix {\bf A} will determine all the fields for $t < t_2
\;$.


\vspace{4ex}

\begin{center}

{\bf 6. Evolution of ${\mathbf e^{\tilde{\lambda}}}$ during the
Planckian regime}

\end{center}

\vspace{2ex}

During the time interval $t_{\tilde{1}} > t > t_2 \;$, the value
of atleast one of the functions $\tilde{f}$ and $f$ remains
maximum $= A \;$. Hence, the universe may be considered to be in
the Planckian regime during this interval. Moreover, during the
sub interval $t_1 > t > t_{\tilde{2}} \;$, one has $\tilde{f} =
f = A$ and $\tilde{\lambda}_t = \lambda_t = 0 \;$, hence
$\tilde{\lambda} = \tilde{\lambda}_1$ and $\lambda = \lambda_1
\;$. Thus, the density $\rho$ remains maximum and the scale
factors $e^{\tilde{\lambda}}$ and $e^\lambda$ remain constant
during this Planckian subperiod.  With no loss of generality, we
take these constant values of the scale factors to be ${\cal O}
(1) \;$, namely take $e^{\tilde{\lambda}_1} \simeq e^{\lambda_1}
\simeq {\cal O} (1) \;$.

Going forward in time from the interval $t_1 > t > t_{\tilde{2}}
\;$, the universe may be considered to be in the classical
regime of general relativity for $t > t_{\tilde{1}}$ when
$\tilde{m}$ and $m$ both lie the interval $(0, \; A)$ and,
hence, $f(x) = x \;$. We will focus on $\tilde{\lambda}$ which,
in an M theory universe considered here, will reach a constant
value in the limit $e^\Lambda \to \infty \;$ in future, causing
the ten dimensional space to become effectively three
dimensional in this limit. During the Planckian regime, as $t$
increases from $t_1$ to $t_{\tilde{1}} \;$, the field
$\tilde{\lambda}$ will evolve and, for the conditions assumed in
equation (\ref{AA}), will increase from $\tilde{\lambda}_1$ to
$\tilde{\lambda}_{\tilde{1}} \;$ whose value may be obtained by
setting $t = t_1$ and $m (t_1) = m_1 = A \;$ in equation
(\ref{tx0}). We have $y_1 = (d - 1) A$ and $y_{\tilde{1}} = (n -
1) m_{\tilde{1}} + \tilde{n} A$ where $m_{\tilde{1}} < A \;$,
see equation (\ref{AA}). Hence, it follows from equations
(\ref{yz1}) and (\ref{rhoy2}), or from equation (\ref{0te1}),
that $\rho (t_1) = \rho_1 = d (d - 1) \; A^2 \; \rho_{qm} \;$,
and then from equation (\ref{tx0}) that
\begin{equation}\label{tx0t11}
e^{(2 \; - \; (d - 1) \; v) \; (\tilde{\lambda}_{\tilde{1}}
- \tilde{\lambda}_1)} \; = \; \frac {\rho_1} {\rho_{\tilde{1}}}
\; = \; \frac {y^2_1 - a^2} {y^2_{\tilde{1}} - a^2} \; \; . 
\end{equation}

The value of the scale factor given above is the result of
Planckian dynamics in our LQC -- inspired model with the
function $f(x)$ given as in equation (\ref{flin}). In the
bi-anisotropic case, the volume of the $\tilde{n}$ dimensional
internal space at time $t_{\tilde{1}}$ is given by
$V_{\tilde{1}} = e^{\tilde{n} \; \tilde{\lambda}_{\tilde{1}}}
\;$. The evolution for $t > t_{\tilde{1}}$ will be as in general
relativity and the internal volume will grow to a constant value
$V_\infty$ as $e^\Lambda \to \infty$, which will occur as $t \to
\infty \;$. In general relativity evolution, with no fine
tuning, $V_\infty \simeq V_{\tilde{1}}$ within a couple of
orders of magnitude \cite{k10}. Hence, a larger value of
$V_{\tilde{1}}$ will result in a larger value of $V_\infty \;$.

We will now estimate the value of the scale factor
$e^{\tilde{\lambda}_{\tilde{1}}} \;$. First consider the factor
$(2 - (d - 1) v) \;$. Let $\tilde{v} = 0 \;$. It then follows
from equation (\ref{tv0v}) that $v = \frac {u} {n - 1} \;$. Note
that, in the Planckian regime, the constituents with lowest $u$
will dominate. Hence, setting $u \simeq 0$ is natural. Setting
$u \simeq \frac {2 (n - 1)} {d - 1}$ can easily result in a
large value for $e^{\tilde{\lambda}_{\tilde{1}}} \;$ but it may
be unphysical in the Planckian regime, see below. 

Now consider the ratio $\frac {\rho_1} {\rho_{\tilde{1}}} \;$
for an M theory universe where $\tilde{n} = 7, \; n = 3$, and $d
= 10 \;$. Note that $\tilde{f} = f_1 = m_1 = A$ and that
\[
\rho_1 = d \; (d - 1) \; A^2 \; \rho_{qm} = 90 \; A^2 \;
\rho_{qm} \; \; . 
\]
It follows from equation (\ref{te1}) that, for $f_{\tilde{1}} =
m_{\tilde{1}} \ge 0 \;$,
\[
\rho_{\tilde{1}} \ge \tilde{n} \; (\tilde{n} - 1) \; A^2 \;
\rho_{qm} = 42 \; A^2 \; \rho_{qm} 
\]
whereas, even for $f_{\tilde{1}} = m_{\tilde{1}} \ge - A \;$,
one only has
\[
\rho_{\tilde{1}} \ge \left( \tilde{n} \; (\tilde{n} - 1)
+ n (n - 1) - 2 n \tilde{n} \right)\; A^2 \; \rho_{qm}
= 6 \; A^2 \; \rho_{qm} \; \; . 
\]
Thus, the ratio $\frac {\rho_1} {\rho_{\tilde{1}}} \le 15$ and,
hence, the scale factor $e^{\tilde{\lambda}_{\tilde{1}}} \;$
increases only by a factor of ${\cal O} (1) $ even though the
universe stays for a long time in the Planckian regime.

Although obtained using a simplified, piece-wise linear
function, it may be that the above results are generic and
indicate that the scale factors may be enhanced by only a factor
of ${\cal O} (1)$ during the Planckian regime in the LQC --
inspired models. However, it is possible that there are other
avenues which may yield larger enhancements. For example : {\bf
(1)} Setting $u \simeq \frac {2 (n - 1)} {d - 1}$ in the
Planckian regime instead of $u \simeq 0$ may be physically
acceptable for some reason, of which we are currently
unaware. In a sense, this would be analogous to an inflation.
Note that when $n = d$, one has $u \simeq 2$ and hence $p = (1 -
u) \rho \simeq - \rho \;$. Naively, one would have expected the
early universe to be dominated by radiation for which $u = 1 -
\frac {1} {d}$ or by matter for which $u$ is even smaller but we
now know that $u \simeq 2$ is physically acceptable under
inflationary conditions. {\bf (2)} In the M theory universe
considered here, we assumed that the densities $\rho_{(*)}$ are
the same for all $* = 2, \; 2', \; 5, \; 5' \;$ in order to
obtain explicit solutions.  Generically, however, these
densities will be different. It is then possible that the total
density may be Planckian, but the constituent densities may
differ sufficiently which may lead to large values for internal
scale factors. Hence a more systematic analysis is needed before
concluding that internal scale factors may be enhanced by only a
factor of ${\cal O} (1)$ during the Planckian regime in the LQC
-- inspired models. Such an analysis, however, is beyond the
scope of the present paper.


\vspace{4ex}

\begin{center}

{\bf 7. Limitations and uses of LQC -- inspired models}

\end{center}

\vspace{2ex}

The LQC -- inspired models generalise empirically the effective
equations in anisotropic LQC, and involve a function $f(x)$ with
the only requirement that $f(x) \to x $ in the limit $x \to 0
\;$. Although the choice of $f(x)$ is otherwise arbitrary, it is
still useful to enquire the genericity of the function used in
this paper and also to enquire, in general, whether the LQC --
inspired models may provide insights into LQC or string/M
theory. Accordingly, we now discuss critically the limitations
and the possible uses of the LQC -- inspired models.

We first consider the limitations. Clearly, these models are not
based on any fundamental principles except that they lead to
general relativity equations in a suitable limit.  Helling
pointed out in \cite{helling} that functions of the form $f(x) =
\sum_n a_n \; sin \; (b_n x)$ should be admissible within the
LQC formalism itself, and argued that a choice of $f$
corresponds to a choice of higher curvature counter terms in the
Einstein -- Hilbert action. But it is not clear to us whether
any choice of $(a_n, b_n)$ is admissible and, if admissible
then, what information about LQC formalism these coefficients
contain. In particular, the functions given in equations
(\ref{fnu}) and (\ref{flin}) may be expressed as above but they
are likely to be non generic and realising them within the LQC
framework, if possible at all, may require special conditions.

Also, the LQC -- inspired models do not generalise all the
effective LQC equations known in different cases but only those
in the anisotropic case. For example, there exist LQC effective
equations for Bianchi type II models \cite{aw2} and type IX
models \cite{w3}. An analysis of these equations suggests that
generalising them empirically will require more functions and,
presently, we are not able to incorporate them in the LQC --
inspired models. Furthermore, even in the isotropic and
anisotropic cases, there are more general LQC effective
equations obtained within the LQG framework. See, for example,
\cite{ydm} -- \cite{gm}. The anisotropic case in these works
require a further generalisation of LQC -- inspired models which
include more functions, see equation (3.7) in a recent paper
\cite{gm} which studies the anisotropic LQC within the LQG
framework.

Thus, clearly, our LQC -- inspired models have many limitations.
However, these models are also useful for several purposes.
They provide a set of equations which give general relativity
equations in a suitable limit. With one arbitrary function
present, these models may be used to study higher dimensional
cosmologiacl evolutions in Planckian regime which are
qualitatively different and, thereby, provide glimpses of Planck
scale physics. Thus, in an earlier work, we have studied a
variety of possible Planckian evolutions and, in this work, we
studied the question of whether large volume of
compactifications are possible with no fine tuning.

In any cosmological evolution which resolves the big bang
singularities, the most interesting questions are the ones about
the observational effects today of the singularity resolutions
and those of the past universe. To deduce such effects, one
needs to know how the past features evolve through the
nonsingular Planckian regime to the present. Hence it is
necessary and important to study in detail the evolution of
cosmological perturbations in non singular universes and to
study their imprints and possible observable consequences. In
LQC, these issues have been studied in great detail using a
variety of methods. See, for example, \cite{fmo} -- \cite{aan3}
which uses `dressed metric approach', \cite{bhks} -- \cite{cbgv}
which uses `closed algebra approach', and \cite{w, w2} which
uses `seperate universe approach', and \cite{am} --
\cite{lzwkcs} for some recent works on this topic.

Clearly, it will be interesting if the cosmological
perturbations and their evolutions may be studied using LQC --
inspired models also. Then the presence of an arbitrary function
in these models may be used to study a variety of possible
observational consequences. However, such a study is beyond the
scope of the present paper but it appears that the dressed
metric approach of \cite{fmo} -- \cite{aan3} may be the
appropriate framework for such studies.

Another use of LQC -- inspired models which may possibly provide
some insight into LQC/G is the following. Dimensional reduction
from $d + 1$ to $n + 1$ dimensional spacetime leads, in general
relativity, to new fields in lower dimensions originating, for
example, from the internal metric components. This can be seen
at the level of the equations of motion also in higher or lower
dimensions. The corresponding structure must also be present in
higher dimensional LQC/G and also in LQC -- inspired models.
However, we are presently unable to disentangle this structure.
If this structure can be found, based on the principle that
effective equations in higher and lower dimensions must be of a
specific form and be transformable into each other, then this
may provide some insight into the structure of higher
dimensional LQC/G.

One may also try to construct an action which leads to the
equations of the LQC -- inspired models. Such an action will
contain higher curvature terms which will depend on the function
$f(x) \;$, see \cite{helling, bo}. If such an action can be
constructed systematically for a given function then, by
comparing it with effective actions in string/M theory, it may
be possible to obtain insights into the later theory.


\vspace{4ex}

\begin{center}

{\bf 8. Conclusion} 

\end{center}

\vspace{2ex}

We now summarise the paper. We studied the evolution of an M
theory universe in the LQC -- inspired models. This universe is
dominated by four stacks of intersecting brane--antibranes and,
in general relativity, it becomes effectively four dimensional
in future while its seven dimensional internal space reaches a
constant size.

In the LQC -- inspired models, we first analysed the conditions
required for non singular evolutions. Then we obtained explicit
solutions by considering a $(\tilde{n} + n)$ dimensional
bi--anisotropic universe where the quantities corresponding to
the $\tilde{n}$ and the $n$ dimensional spaces are seperately
isotropic, and by considering a simplified, piece--wise linear
function for which the evolutions are non singular.

We applied these solutions to the M theory universe and
considered the question of whether the physics in the non
singular Planckian regime can enhance the future constant size
of its seven dimensional internal space. Using the explicit
solutions, we found no non trivial enhancement of this size.
This may be a generic feature of the LQC -- inspired models but
it is also possible that there are other avenues which may yield
larger enhancements.

We have also discussed critically the limitations and the uses
of our models. We now conclude by mentioning a few topics for
further studies where we think that some progress may be
possible in the near future. The LQC -- inspired models involve
a function, the choice of which leads to a variety of
evolutions. It is desireable to understand the origin of this
function and to explore the physical principles which may
restrict it as uniquely as possible.

In string/M theory, effective higher derivative actions can be
constructed systematically. It is worthwhile to explore whether
the equations of motion resulting from these higher derivative
actions bear any relation to the effective equations in LQC or
to the equations in the LQC -- inspired models.

The M theory considered here becomes effectively four
dimensional in future while its seven dimensional internal space
reaches a constant size. Its evolution can be made non singular
in the LQC -- inspired models. In such a set up, one can now
explore various mechanisms which may lead to large internal
volumes which are of phenomenological interest \cite{add,
cq}. It will be equally interesting if one can prove instead,
either in general or within the LQC -- inspired models, that
such large internal volumes are not possible.


\vspace{4ex}

{\bf Acknowledgement:} We thank the referee for helpful comments.


\vspace{4ex}

\begin{center}

{\bf Appendix A : Anisotropic solutions in general relativity}

\end{center}

\vspace{2ex}

Consider the general relativity equations (\ref{t21}) --
(\ref{rhot}) for the anisotropic case. When the equations of
state are linear, it is straightforward to solve these equations
and obtain analytic solutions \cite{k10}. It follows from
equations (\ref{gic}) that, upon replacing $\kappa^2$ by $c^2
\kappa^2 \;$, these solutions are applicable to the LQC --
inspired models when $f(x) = c x + c_0 \;$. 

We now present these solutions. First, define a new variable
$\tau$ by
\begin{equation}\label{dtau}
d t = e^\Lambda \; d \tau
\; \; \; \longleftrightarrow \; \; \; 
t - t_0 = \int_{\tau_0}^\tau d \tau \; e^\Lambda 
\end{equation}
where $t_0$ and $\tau_0$ are initial times. Then, for any
function $\psi (t(\tau)) \;$, we have
\[
\psi_\tau \; = \; e^\Lambda \; \psi_t \; \; , \; \; \; 
\psi_{\tau \tau} \; = \; e^{2 \Lambda} \; (\psi_{t t}
+ \Lambda_t \psi_t) \; \; .
\]
Defining $(\hat{*}) = e^{2 \Lambda} \; (*)$ for $ (*) = (\rho,
\; p_i, \;r^i) \;$, equations (\ref{t21}) -- (\ref{rhot}) become
\begin{eqnarray} 
\sum_{i j} G_{i j} \; \lambda^i_\tau \; \lambda^j_\tau
& = & 2 \kappa^2 \; \hat{\rho} \label{hat21} \\
& & \nonumber \\
\lambda^i_{\tau \tau} & = & \kappa^2 \; \hat{r}^i
\label{hat22} \\
& & \nonumber \\
(\hat{\rho})_\tau & = & \sum_i (\hat{\rho} - \hat{p}_i)
\; \lambda^i_\tau  \; \; . \label{hatrhot}
\end{eqnarray}
Let the equations of state be linear and be given by
\begin{equation}\label{ui}
p_i = (1 - u_i) \; \rho
\end{equation} 
where $u_i$ are constants. Define $l, \; v^i$, and ${\cal G}$ by
\begin{equation}\label{lv}
l = \sum_i u_i \; \lambda^i \; \; , \; \; \; 
v^i = \sum_j G^{i j} \; u_j \; \; , \; \; \; 
{\cal G} = \sum_i v^i \; u_i = \sum_{i j} G^{i j} \; u_i \; u_j
\end{equation}
and let the initial values of various quantities at $t = t_0$ be
given by
\begin{eqnarray}
& \left( \lambda^i, \; \lambda^i_t, \; \rho \; ; \; \Lambda,
\; l, \; l_t \; ; \; \tau, \; \lambda^i_\tau, \; l_\tau,
\; \hat{\rho} \right)_{t = t_0} & \nonumber \\
& & \nonumber \\
= & \left( \lambda^i_0, \; k^i, \; \rho_0\; ; \; \Lambda_0,
\; l_0, \; l_{t 0} \; ; \; \tau_0, \; \lambda^i_{\tau 0}, \;
l_{\tau 0}, \; \hat{\rho}_0 \right) & \label{ic}
\end{eqnarray}
where 
\begin{eqnarray}
\rho_0 \; > \; 0 & , & \sum_{i j} G_{i j} \; k^i \; k^j
\; = \; 2 \kappa^2 \; \rho_0 \nonumber \\
& & \nonumber \\
\Lambda_0 = \sum_i \lambda^i_0 & , &
l_0 = \sum_i u_i \; \lambda^i_0 \; \; , \; \; \; 
l_{t 0} = \sum_i u_i \; k^i \nonumber \\
& & \nonumber \\
\lambda^i_{\tau 0} = e^{\Lambda_0} \; k^i & , & 
l_{\tau 0} = e^{\Lambda_0} \; l_{t 0} \; \; , \; \; \;
\hat{\rho}_0 = e^{2 \Lambda_0} \; \rho_0
\; \; . \label{icrelns}
\end{eqnarray}
Then equations (\ref{hat22}) and (\ref{hatrhot}) give
\begin{eqnarray}
\lambda^i_{\tau \tau} & = & \kappa^2 \; v^i \; \hat{\rho}
\label{hat32} \\
& & \nonumber \\
l_{\tau \tau} & = & \kappa^2 \; {\cal G} \; \hat{\rho}
\label{lItt} \\
& & \nonumber \\
\hat{\rho} & = & \hat{\rho}_0 \; e^{l - l_0} 
\label{rhol} 
\end{eqnarray}
and it follows from equations (\ref{hat32}) and (\ref{lItt})
that
\begin{equation}\label{lambdai} 
\lambda^i - \lambda^i_0 \; = \; \frac {v^i} {\cal G} \;
(l - l_0) + L^i \; (\tau - \tau_0) \; \; . 
\end{equation}
Since $l = \sum_i u_i \lambda^i \;$, it follows that the
integration constants $L^i$ must satisfy the constraint $\sum_i
u_i L^i = 0 \;$. This constraint is identically satisfied if
$L^i = e^{\Lambda_0} \left( k^i - \frac {v^i} {\cal G} \; \;
l_{t 0} \right)$ where $l_{t 0} = \sum_i u_i k^i \;$, see
equations (\ref{icrelns}). Thus, the set of $d$ number of
initial values $\{ k^i \}$ is equivalent to the set of $(1 + d)$
number of initial values $\{ l_{t 0}, \; L^i \}$ together with
one constraints on $L^i \;$. Upon using $\sum_i u_i L^i = 0 \;$,
equation (\ref{hat21}) gives
\begin{equation}\label{l2E} 
(l_\tau)^2 \; = \; 2 \; {\cal G} \; \left(E
+ \kappa^2 \; \hat{\rho} \right) \; \; , \; \; \;
2 \; E = - \; \sum_{i j} G_{i j} \; L^i \; L^j \; \; . 
\end{equation} 
Now, in principle, equations (\ref{lItt}), (\ref{rhol}), and
(\ref{l2E}) give $l(\tau) \;$ and equations (\ref{lambdai}) and
(\ref{dtau}) give $\lambda^i (\tau)$ and $t(\tau)$ from which
$\tau (t), \; l (t)$, and $\lambda^i (t)$ follow. Also, it can
be shown that if $\sum_i u_i L^i = 0 \;$ and ${\cal G} = \sum_{i
j} G^{i j} \; u_i u_j > 0$ then $E \ge 0 \;$ and $E$ will vanish
if and only if all $L^i$ vanish \cite{k10}. Henceforth, we
assume that ${\cal G} > 0$ and $E > 0 \;$.

For the case of bi--anisotropic universe considered in this
paper, see equations (\ref{lqtll}) and (\ref{dstll}), it follows
straightforwardly that
\begin{eqnarray}
\tilde{p} = (1 - \tilde{u}) \; \rho & , &
p = (1 - u) \; \rho \nonumber \\
& & \nonumber \\
\tilde{v} = \frac {n u - (n - 1) \tilde{u}}
{\tilde{n} + n - 1} & , &
v = \frac {\tilde{n} \tilde{u} - (\tilde{n} - 1)
u} {\tilde{n} + n - 1} \nonumber \\
& & \nonumber \\
l = \tilde{n} \tilde{u} \tilde{\lambda} + n u \lambda
& , & {\cal G} = \tilde{n} \tilde{u} \tilde{v}
+ n u v \nonumber \\
& & \nonumber \\
\tilde{n} \tilde{u} \tilde{L} + n u L = 0 & , &
2 E = \frac {n (n + \tilde{n} - 1) {\cal G} L^2} {\tilde{n}
\tilde{u}^2}  \label{L1E}
\end{eqnarray} 
where the expression for $E$ follows after some algebra. If
$\tilde{v} = 0 \;$, which is necessary for the $(\tilde{n} + n)$
dimensional space to become effectively $n$ dimensional in the
limit $e^\Lambda \to \infty \;$, then one has
\begin{eqnarray}
\tilde{u} = \frac {n u} {n - 1} & , &
v = \frac {u} {n - 1} \nonumber \\
& & \nonumber \\
l = \tilde{u} \; \left( \tilde{n} \tilde{\lambda}
+ (n - 1) \lambda \right)
& , & {\cal G} = \frac {n u^2} {n - 1} \nonumber \\
& & \nonumber \\
\tilde{n} \tilde{L} + (n - 1) L = 0 & , & 2 E = \frac
{(n - 1) (n + \tilde{n} - 1) L^2} {\tilde{n}} \; \; . \label{LE}
\end{eqnarray}

Consider now the solution $l(\tau)$ for equations (\ref{lItt}),
(\ref{rhol}), and (\ref{l2E}). As can be verified easily, it is
given by
\begin{equation}\label{lsoln} 
\kappa^2 \; \hat{\rho} \; = \; 
\kappa^2 \; \hat{\rho}_0 \; e^{l - l_0} \; = \;
\frac {E} {Sinh^2 \; \sigma (\tau_\infty - \tau)}
\end{equation}
where $2 \sigma^2 = {\cal G} E \;$. Note that the sign of
$\sigma$ is immaterial; that
\begin{equation}\label{ltausoln}
l_\tau \; = \; 2 \; \sigma \;
Coth \; \sigma (\tau_\infty - \tau) \; \; ; 
\end{equation}
and that setting $l = l_0$ and $\tau = \tau_0$ in equation
(\ref{lsoln}) gives $\tau_\infty$ in terms of $E$ and
$\hat{\rho}_0 \;$. Equations (\ref{lambdai}) and (\ref{dtau})
will now give $\lambda^i (\tau)$ and $t(\tau)$ from which $\tau
(t), \; l (t)$, and $\lambda^i (t)$ follow. Taking $\sigma > 0$
and $l_{\tau 0} > 0$ for the sake of definiteness, we now
mention some features of these solutions.


\begin{itemize}

\item

Since $l_{\tau 0} > 0$, it follows from equation
(\ref{ltausoln}) that $\tau_\infty > \tau_0 \;$. It follows from
equation (\ref{lsoln}) that $l(\tau)$ varies monotonically
between $- \infty$ and $+ \infty$, that $l \to - \infty$ as
$\tau \; \to \; - \infty \;$, and that $l \to \infty$ as $\tau
\; \to \; \tau_\infty \;$.

\item

In the limit $\tau \; \to \; \tau_\infty$ from below, one has
$l(\tau) \; \sim \; - 2 \; ln \; (\tau_\infty - \tau) \; \to \;
\infty \;$. Equations (\ref{lambdai}) and (\ref{dtau}) then
give, upto unimportant constants,
\begin{eqnarray}
t & \sim & (\tau_\infty - \tau)^{- \frac {2 B - {\cal G}}
{\cal G}} \; \; , \; \; \; B = \sum_j v^j \label{ttau} \\
& & \nonumber \\
e^{\lambda^i} & \sim & (\tau_\infty - \tau)^{- \frac {2 v^i}
{\cal G}} \; \sim \; t^{\frac {2 v^i} {2 B - {\cal G}}}
\label{littau} \\
& & \nonumber \\
e^\Lambda & \sim & (\tau_\infty - \tau)^{- \frac {2 B} {\cal G}}
\; \sim \; t^{\frac {2 B} {2 B - {\cal G}}} \label{lttau}
\; \; .
\end{eqnarray} 
For the bi--anisotropic universe, it follows from equations
(\ref{L1E}) that
\begin{equation}\label{2b-g}
2 B - {\cal G} = \tilde{n} \tilde{v} \;
(2 - \tilde{u}) + n v \; (2 - u) \; \; . 
\end{equation}
Hence, for $\tilde{v} = 0 \;$, one has $e^{ \tilde{\lambda}}
\sim const \;$ and $e^{\lambda} \sim t^{\frac {2} {n (2 - u)}}
\;$ which is the standard $n$ dimensional result.

\item 

In the limit $\tau \; \to \; - \infty \;$, one has $l(\tau) \;
\sim \; 2 \; \sigma \tau \; \to \; - \infty \;$. Equations
(\ref{lambdai}) then imply that $\lambda^i (\tau)$ are all
linear in $\tau \;$. Let $\tau \to - \infty$ and $e^\Lambda \to
0$ in this limit and, upto unimportant constants, let
\[
\lambda^i \; \sim \; q^i \; \tau \; \; , \; \; \;
\Lambda \; \sim \; q \; \tau \; \; , \; \; \; 
q = \sum_i q^i > 0 \; \; . 
\]
Then, after some algebra, it follows from equation (\ref{dtau})
that
\begin{equation}\label{lli} 
e^\Lambda \; \sim \; e^{q \; \tau} \; \sim \; q \; t
\; \; \to \; 0 \; \; \; \; , \; \; \; \; \;
e^{\lambda^i} \; \sim \; e^{q^i \; \tau} \; \sim \;
\left(q \; t \right)^{\frac {q^i} {q}} 
\end{equation}
which are the Kasner--type solutions.

\end{itemize}


\vspace{4ex}

\begin{center}

{\bf Appendix B : Isotropic solutions in LQC -- inspired models}

\end{center}

\vspace{2ex}

Consider the fully isotropic case where 
\[
(m^i, \; f^i, \; g_i, \; X_i, \; \lambda^i, \; p_i, \; r^i )
\; = \; (m, \; f, \; g, \; X, \; \lambda, \; p, \; r ) 
\] 
for $i = 1, 2, \cdots, d \;$. Then
\[
g \; = \; \frac{d \; f} {d m} \; \; , \; \; \;
X \; = \; (d - 1) \; g f \; \; , \; \; \;
r \; = \; \frac {\rho - p} {d - 1} 
\]
and equations (\ref{e1}) -- (\ref{e3}) give
\begin{eqnarray}
f^2 & = & \frac {2 \; \gamma^2 \lambda_{qm}^2 \kappa^2 \;
\rho} {d \; (d - 1)} \label{eiso1} \\
& & \nonumber \\
m_t & = & - \; \frac {\gamma \lambda_{qm} \kappa^2} {d - 1} \;
(\rho + p) \label{eiso2} \\
& & \nonumber \\ 
\lambda_t & = & \frac {g \; f} {\gamma \lambda_{qm}}
\; \; \; \Longrightarrow \; \; \; 
(\lambda_t)^2 \; = \; \frac {2 \kappa^2 \; (\rho \; g^2)}
{d (d - 1)} \; \; . \label{eiso3}
\end{eqnarray}
Let the equation of state be linear and be given by $p = (1 - u)
\rho$ where $u < 2$ is a constant. Then equations (\ref{rhot})
and (\ref{eiso1}) -- (\ref{eiso3}) may be solved explicitly if
certain integrations and functional inversions can be
performed. Equations (\ref{rhot}) and (\ref{eiso1}) give
\begin{equation}\label{am} 
\frac {\rho} {\rho_0} \; = \; \frac {f^2} {f^2_0}\; = \; 
e^{ - (2 - u) \; d \; (\lambda - \lambda_0)}
\end{equation}
which leads to $\lambda(m) \;$. Equations (\ref{eiso1}) and
(\ref{eiso2}) then lead to $t(m)$ given by
\begin{equation}\label{tm}
c_{qm} \; (t - t_0) \; = \; - \; \int^m_{m_0} \frac {d m} {f^2}
\end{equation}
where $c_{qm} = \frac {(2 - u) \; d} {2 \; \gamma \lambda_{qm}}
\;$. Inverting $t(m)$ then gives $m (t)$ and $\lambda (t)
\;$. The integrations and functional inversions required here
can be performed explicitly for $f(x) = c x + c_0$ and also for
$f(x) = sin \; x \;$ but not for a generic $f(x) \;$. The
resulting solutions are given in \cite{k16, k17}.

\vspace{2ex}

Consider now the isotropic solutions for the simplified,
piece-wise linear function $f(x)$ given in equation
(\ref{flin}). Equation (\ref{am}) gives the density $\rho (m)$
and the scale factor $e^{\lambda (m)} \;$.  Let the initial
value $m_0$ at time $t_0$ lie in the range $0 < m_0 < A \;$. It
then follows that as $m$ increases from $0$ to $m_0$ to $A$ to
$A + 2 \Delta$ to $2 m_*$, the function $f$ increases from $0$
to $m_0$ to $A$, remaining at $A$, and then decreasing to $0
\;$. Hence, correspondingly, the scale factor $e^{\lambda (m)}$
decreases from $\infty$ to $e^{\lambda_0}$, decreases further,
then remains constant, and then increases again to $\infty \;$.

The time $t(m)$ follows straightforwardly upon performing the
integration in equation (\ref{tm}), and is given by
\begin{eqnarray}
c_{qm} \; (t - t_0) \; = \; - \frac {1} {m_0} + \frac {1} {m}
\; \; \; & for & \; \; 0 \le m \le A \nonumber \\
& & \nonumber \\
\; = \; \frac {2} {A} - \frac {1} {m_0} - \frac {m} {A^2}
\; \; \; & for & \; \; A \le m \le A + 2 \Delta
\nonumber \\ 
& & \nonumber \\
\; = \; \frac {2} {A} - \frac {1} {m_0} - \frac {2 \Delta} {A^2}
+ \frac {1} {m - 2 m_*}
\; \; \; & for & \; \; A + 2 \Delta \le m \le 2 m_*
\; \; . \label{tmlin}
\end{eqnarray}
Hence, as $m$ increases from $0$ to $m_0$ to $A$ to $A + 2
\Delta$ to $2 m_* \;$, the time $t$ decreases monotonically from
$\infty$ to $t_0$ to $- \infty$, first as $\frac {1} {m} \;$,
then linearly as $- \frac {m} {A^2} \;$, and then as $\frac {1}
{m - 2 m_*} \;$.


\vspace{4ex}

\begin{center}

{\bf Appendix C : Bi--anisotropic solutions  

\vspace{2ex}

when only $\mathbf{ m \; \in \; (A, \; A + 2 \Delta)}$}

\end{center}

\vspace{2ex} 

During $t_b > t > t_e \;$, let $m (t)$ lie in the interval $(A,
\; A + 2 \Delta)$ and let $\tilde{m} (t)$ lie in $(0, \; A)$ or
in $(A + 2 \Delta, \; 2 m_*) \;$. Then $f = A$, $\; g = X = 0$,
$\; \tilde{f} = c \tilde{m} + c_0$ where $(c, \; c_0) = (1, \;
0)$ or $(- 1, \; 2 m_*)$, $\; \tilde{g} = c \;$, and $\tilde{X}
= c \left( (\tilde{n} - 1) \tilde{f} + n A \right) \;$. The
times $t_b$ and $t_e$ are defined by the equalities in the
following expressions for the values of $\tilde{m}$ and $m$ at
$t_b$ and $t_e \;$ :
\begin{eqnarray}
\tilde{m}_b \; < \; A & , &  m_b \; = \; A
\nonumber \\
& & \nonumber \\
\tilde{m}_e \; = \; A & , & A \; < \; m_e \; < \; A + 2 \Delta
\nonumber \\
& & \nonumber \\
or \; \; \;  \; \; \; 
\tilde{m}_b \; = \; A + 2 \Delta & , &
A \; < \; m_b \; < \; A + 2 \Delta 
\nonumber \\
& & \nonumber \\
\tilde{m}_e \; > \; A + 2 \Delta & , & m_e \; = \; A + 2 \Delta
\; \; . \label{cAA}
\end{eqnarray}
Several equations are different if $\tilde{n} > 1$ or $\tilde{n}
= 1 \;$. Hence we consider these two cases seperately.


\vspace{4ex}

\centerline{ $\mathbf \tilde{n} > 1 \;$ }

\vspace{2ex}

Define $y, \; z$, and $a$ by
\begin{equation}\label{cyz2}
y = (\tilde{n} - 1) \; \tilde{f} + n \; A \; \; , \; \; \;
z = (m - \tilde{m}) \; \; , \; \; \; 
a = \sqrt{\frac {n \; (d - 1)} {\tilde{n}}} \; A \; \; .
\end{equation}
Note that $\tilde{X} = c y \;$ and that $n A < y < (d - 1) A
\;$. After some algebra, it follows from equations (\ref{te1})
and (\ref{te2}) that
\begin{eqnarray}
\frac {\rho} {\rho_{qm}} & = &
\frac {\tilde{n}} {\tilde{n} - 1} \left( y^2 - a^2 \right)
\label{crhoy2} \\
& & \nonumber \\
y_t & = & - \; c_y \; (y^2 - a^2) \label{2yt} \\
& & \nonumber \\
z_t + b \; y \; z & = & - \; c_z \; (y^2 - a^2) \label{2zt}
\end{eqnarray} 
where 
\[
c_y \; = \; \frac {\tilde{n} \; c \; \left( \frac {2} {d - 1} -
\tilde{v} \right)} {2 \; \gamma \lambda_{qm}}
\; \; , \; \; \; 
c_z \; = \; \frac {\tilde{n} \; (\tilde{v} - v)} {2 \;
(\tilde{n} - 1) \; \gamma \lambda_{qm}}
\; \; , \; \; \; 
b \; = \; \frac {\tilde{n} \; c} {(d - 1) \; \gamma
\lambda_{qm}} \; \; . 
\]
The solutions $y(t)$ and $z(y)$ are given in equations
(\ref{ysoln}) and (\ref{zsoln}). Since $X = 0$, it follows that
$\Lambda_t -\lambda_t = 0$ and hence, from equations
(\ref{tvv}), (\ref{rhotll}), and (\ref{crhoy2}), that
\begin{eqnarray}
(\tilde{\lambda} - \tilde{\lambda}_0) & = & - \; \left( \frac
{n - 1} {\tilde{n}} \right) \; (\lambda - \lambda_0)
\nonumber \\
& & \nonumber \\
e^{(2 \; - \; (d - 1) \; \tilde{v}) \; (\lambda_b - \lambda)}
& = & \frac {\rho} {\rho_b} \; = \;
\frac {y^2 - a^2} {y^2_b - a^2} \; \; . \label{cx0}
\end{eqnarray}


\vspace{4ex}

\centerline{ $\mathbf \tilde{n} = 1 \;$ }

\vspace{2ex}

Now $d = n + 1 \;$. Define $y$ and $z$ by
\begin{equation}\label{cyz21}
y = 2 \; \tilde{f} + (n - 1) \; A \; \; , \; \; \;
z = (m - \tilde{m}) \; \; . 
\end{equation}
Note that $\; \tilde{X} = n c A \;$ and that $(n - 1) A < y < (n
+ 1) A \;$. After some algebra, it follows from equations
(\ref{te1}) -- (\ref{te2}) that
\begin{eqnarray}
\frac {\rho} {\rho_{qm}} & = & n A \; y
\label{c21te1} \\
& & \nonumber \\
y_t & = & 2 c \; \gamma \lambda_{qm} \kappa^2 \; \left(
\tilde{v} - \frac {2} {d - 1} \right) \; \rho
\label{c21yt} \\
& & \nonumber \\
z_t \; + \; \frac {n c \; A \; z}
{(d - 1) \; \gamma \lambda_{qm}} & = & \gamma \lambda_{qm}
\kappa^2 \; (v - \tilde{v}) \; \rho \; \; . \label{c21zt}
\end{eqnarray}
Equations (\ref{c21te1}) -- (\ref{c21zt}) lead to the solutions
$y(t)$ and $z(y)$ given by
\begin{equation}\label{cy1soln}
y \; = \; y_b \; e^{ - \frac {n c \; A } { \gamma \lambda_{qm}}
\; \left( \frac {2} {d - 1} - \tilde{v} \right) \; (t - t_b)}
\end{equation}
and
\begin{equation}\label{cz1soln}
z \; = \; y^s \; \left( \frac {z_b} {y^s_b}
\; + \; \sigma \; 
\; \int_{y_b}^y \; \frac {d y} {y^s} \right) \; \; .
\end{equation}
where $s = \frac {1} {2 - (d - 1) \; \tilde{v}} \;$ and $\sigma
= \frac {(d - 1) \; (\tilde{v} - v)} {2 c \; \left( 2 - (d - 1)
\; \tilde{v} \right)} \;$. Thus, if $\tilde{v} < \frac {2} {d -
1}$ then $y_t < 0 \;$ and $y$ increases monotonically from $y_b$
to $\infty$ as $t$ decreases from $t_b$ to $ - \infty \;$. Also,
equations (\ref{cx0}) give $\tilde{\lambda}$ and $\lambda$ in
terms of $y \;$.


\end{document}